\pdfoutput=1

\documentclass[
    superscriptaddress,
    amsmath,amssymb,
    aps,
    nofootinbib,
]{revtex4-2}

\usepackage{graphicx}
\usepackage{bm}
\usepackage[svgnames]{xcolor}
\usepackage[colorlinks,citecolor=DarkGreen,linkcolor=FireBrick,linktocpage,unicode]{hyperref}
\usepackage{mathtools}
\usepackage{booktabs,tabularx, makecell, array}  
\usepackage{orcidlink}
\usepackage{longtable}

\newcommand{\br}{{\bm r}}
\newcommand{\be}{{\bm e}}

\newcommand{\cH}{\mathcal{H}}
\newcommand{\cO}{\mathcal{O}}

\newcommand{\bp}{{\bm p}}
\newcommand{\bF}{{\bm F}}
\newcommand{\bI}{{\bm I}}
\newcommand{\bh}{{\bm h}}
\newcommand{\bP}{{\bm P}}
\newcommand{\bK}{{\bm K}}
\newcommand{\bU}{{\bm U}}
\newcommand{\bx}{{\bm x}}
\newcommand{\by}{{\bm y}}
\newcommand{\bu}{{\bm u}}
\newcommand{\pg}{{\mathbf{p}_g}}
\newcommand{\Gg}{{\mathbf{G}_g}}
\newcommand{\Tr}{\mathrm{Tr}}
\newcommand{\exprel}{\mathrm{exprel}}
\newcommand{\pd}[2]{\frac{\partial #1}{\partial #2}}

\begin{document}

\preprint{perspective}

\title{Conserved quantities and ensemble measure for Martyna--Tobias--Klein barostats with restricted cell degrees of freedom}

\author{Kohei Shinohara  \orcidlink{0000-0002-5907-2549}} 
    \thanks{This work was conducted independently and does not reflect the views or interests of the author's employer.}
    \email{kshinohara0508@gmail.com}
    \affiliation{Preferred Networks, Inc., Tokyo 100-0004, Japan.}

\date{\today}%


\begin{abstract}
We derive the conserved energy-like quantity and ensemble measure for Martyna--Tobias--Klein (MTK) barostats in which only a restricted subset of the cell degrees of freedom are active.
In the standard fully anisotropic MTK formulation the number of barostat degrees of freedom is $d^{2}$, where $d$ is the spatial dimension.
When only $n_c$ axes of the cell matrix are allowed to fluctuate, the conserved energy-like quantity retains the same functional form but with $d^{2}$ replaced by $n_c$ in every term that counts barostat degrees of freedom.
The derivation builds on the generalized Liouville framework for non-Hamiltonian systems and the existing MTK integration machinery.
We verify that this quantity is exactly conserved, show that the resulting dynamics samples the isothermal--isobaric ensemble restricted to the submanifold of cell shapes in which inactive components are held fixed, and provide a complete Liouville-operator-based integration scheme for the masked MTK variant.
\end{abstract}

\maketitle

\section{\label{sec:intro}Introduction}

The Martyna--Tobias--Klein (MTK) equations of motion~\cite{Martyna1994} extend the Nos\'e--Hoover chain (NHC) thermostat~\cite{Martyna1992,Tuckerman2010} to constant-pressure molecular dynamics by coupling the simulation cell to a barostat variable.
When combined with a measure-preserving, time-reversible integrator derived from the Liouville-operator formalism~\cite{Tuckerman2006}, the MTK scheme provides a robust route to sampling the isothermal--isobaric ($NPT$) ensemble.

In many practical applications, one does not wish to allow the full cell tensor to fluctuate.
For example, slab geometries call for pressure control only along the surface normal ($NP_{z}T$ ensemble), while uniaxial-stress simulations constrain the lateral dimensions.
Codes such as LAMMPS~\cite{Thompson2022} already support such restricted-axis ensembles, following the Shinoda formulation~\cite{Shinoda2004}.
The original MTK paper~\cite{Martyna1994} derives the conserved energy-like quantity for the isotropic and fully anisotropic ($d^{2}$ degrees of freedom) cases, but does not treat the intermediate case in which a subset of cell axes is active.
To the best of my knowledge, no compact published derivation exists for the corresponding reduced-degree-of-freedom conserved energy-like quantity and ensemble measure.

Two restrictions apply to the present formulation.
First, we address only diagonal cell-length fluctuations; the extension to off-diagonal tilt factors is left for future work.
Second, each active cell axis must be orthogonal to all other axes (Sec.~\ref{sec:masked}), which excludes certain non-orthogonal cell shapes from partial barostatting.
This orthogonality condition is automatically satisfied for orthorhombic cells.
For a hexagonal cell, the unique six-fold axis is orthogonal to the basal plane and may be barostatted alone, but simultaneous control of both in-plane basis vectors requires reformulation in an orthohexagonal supercell because they are not orthogonal.
A fully triclinic cell cannot satisfy the condition at all.

This short note fills the derivation gap identified above: we recall the generalized Liouville framework for non-Hamiltonian systems (Sec.~\ref{sec:nonham}), derive the equations of motion, the conserved energy-like quantity, and the expected ensemble for the masked MTK barostat (Sec.~\ref{sec:masked}), and provide the Liouville-operator-based integration scheme for the masked variant (Appendix~\ref{app:masked-integrator}).
For completeness, Appendices~\ref{app:nhc}--\ref{app:aniso} summarize the standard integration schemes---the NHC thermostat, the isotropic MTK barostat, and the fully anisotropic MTK barostat---on which Appendix~\ref{app:masked-integrator} builds.
Sec.~\ref{sec:nonham} and Appendices~\ref{app:nhc}--\ref{app:aniso} review established material from Refs.~\cite{Martyna1992,Martyna1994,Tuckerman2006,Tuckerman2010}; the novel contributions of this note are contained in Sec.~\ref{sec:masked} and Appendix~\ref{app:masked-integrator}.

The implementation of the masked MTK barostat in \textsc{ASE}~\cite{ase-paper} is available at \href{https://ase-lib.org/ase/md.html#}{\texttt{ase.md.nose\_hoover\_chain.MaskedMTKNPT}}.


\section{\label{sec:nonham}Non-Hamiltonian systems}

The MTK equations of motion are non-Hamiltonian: the thermostat and barostat couplings break the symplectic structure.
Their statistical-mechanical validity rests on the generalized Liouville framework for non-Hamiltonian systems~\cite{Tuckerman_1999,10.1063/1.1378321,Tuckerman2006,Tuckerman2010}, which we now summarize.

We consider a dynamical system
\begin{align}
  \dot{\bx} = \boldsymbol{\xi}(\bx)
\end{align}
for $n$-dimensional phase-space variables $\bx$.
We write an initial condition $\bx_0$ and the corresponding trajectory $\bx_t := \bx(\bx_0, t)$ at time $t$.
The Liouville operator
\begin{align}
  i \mathcal{L} &\coloneqq \boldsymbol{\xi}(\bx) \cdot \nabla_{\bx},
  \label{eq:liouville-op}
\end{align}
gives the formal solution $\bx_t = e^{i \mathcal{L} t}\, \bx_0$.
Factorizing $e^{i \mathcal{L} \Delta t}$ via symmetric Trotter splittings yields time-reversible, measure-preserving integrators~\cite{Tuckerman2006}.

The phase-space compressibility of the dynamical system is defined as
\begin{align}
  \kappa(\bx)
    &\coloneqq \nabla_{\bx} \cdot \dot{\bx} = \sum_{i=1}^n \pd{\xi_i (\bx)}{x_i}.
\end{align}
The phase-space volume elements $d\bx_0$ and $d\bx_t$ are related by the phase-space compressibility as
\begin{align}
  d\bx_t = \exp \left( \int_{0}^{t} \kappa(\bx_s) ds \right) d\bx_0.
\end{align}
The measure $e^{-w(t)} d\bx_t$ is conserved along the trajectory, where $w(t)$ is an indefinite time integral of $\kappa(\bx)$ along the trajectory,
\begin{align}
  w(t) &\coloneqq \int^{t} \kappa(\bx_s) ds.
\end{align}
Note that we drop the dependence of $w(t)$ on the initial condition $\bx_0$ when it is clear from context.

We suppose that the system has $m$ independent conserved quantities $\Lambda_k (\bx) \, (k = 1, \dots, m)$.
For the initial condition $\bx_0$, we write the values as $C_k = \Lambda_k (\bx_t)$.
The microcanonical partition function of this system with conserved quantities $\{ C_k \}$ is defined as
\begin{align}
  \Omega(\{ C_k \}) = \int d\bx\, e^{-w(t)} \prod_{k=1}^m \delta(\Lambda_k (\bx) - C_k).
\end{align}
Because the measure $e^{-w} d\bx$ is conserved along the trajectory, we can sample the microcanonical distribution by running a single trajectory and accumulating time averages with the assumption of ergodicity.

\section{Masked MTK barostat}
\label{sec:masked}

Throughout this section, $d$ denotes the spatial dimension, $N$ the number of particles, and $N_f = dN$ the number of physical degrees of freedom (in the absence of constraints).
We write $\br_i$, $\bp_i$, $m_i$, and $\bF_i$ for the position, momentum, mass, and force on particle~$i$, and $\bh$ for the $d \times d$ cell matrix whose rows are the basis vectors.

We consider the case where only $n_c$ of the $d$ cell axes are allowed to fluctuate, i.e.\ the barostat momentum $\pg$ is restricted to have at most $n_c$ nonzero diagonal entries while the cell matrix $\bh$ itself may have arbitrary shape.
Label the first $n_c$ axes as active and the remaining $d - n_c$ axes as inactive, and let $\be_k$ be the unit vector along the $k$-th cartesian axis associated with the cell matrix $\bh$, not a fixed cartesian unit vector.
We denote the length of the $k$-th axis as $\lambda_k$,
\begin{align}
  \bh &= \begin{pmatrix}
    \lambda_1 \be_1^\top \\
    \vdots \\
    \lambda_{d} \be_{d}^\top
  \end{pmatrix}.
\end{align}
The masked barostat momentum is restricted to the subspace spanned by the active axes:
\begin{align}
  \pg = \sum_{c=1}^{n_c} p_c\, \be_c \be_c^{\top},
  \label{eq:pg-masked}
\end{align}
where $p_c$ is the scalar momentum conjugate to axis $c$.
We assume the active axes are orthogonal to all other axes,
\begin{align}
  \be_c^\top \be_k = 0, \quad c = 1, \dots, n_c,\; k \neq c.
  \label{eq:ortho-assumption}
\end{align}
This condition is automatically satisfied for orthorhombic cells.
For non-orthogonal cells it restricts which axes may be active: in a hexagonal cell the unique six-fold axis is orthogonal to the basal plane and may be barostatted alone, but activating both in-plane hexagonal basis vectors would violate orthogonality because they are not orthogonal; one must instead adopt an orthohexagonal supercell.
A fully triclinic cell, in which no pair of axes is in general orthogonal, falls outside the scope of this formulation entirely.

\subsection{\label{sec:eom}Equations of motion}

To make the masked formulation explicit, we first write the physical Hamiltonian and instantaneous internal stress tensor:
\begin{align}
  \cH(\br, \bp) &\coloneqq \sum_{i=1}^{N} \frac{\bp_i^{2}}{2m_i} + U(\br), \\
  \mathbf{P}^{\mathrm{int}}
  &= \frac{1}{\det \bh}
    \sum_{i=1}^{N}
    \left[
      \frac{\bp_{i} \bp_{i}^\top}{m_i}
      + \bF_{i} \br_{i}^\top
    \right],
  \label{eq:stress-masked}
\end{align}
where we assume the potential energy $U$ depends only on the particle positions $\br$ and not on the cell $\bh$.
The thermostat chain variables are $(\eta_j, p_{\eta_j})$ and the barostat chain variables are $(\xi_j, p_{\xi_j})$ for $j = 1, \dots, M$.
Their driving forces are
\begin{subequations}
  \label{eq:masked-driving}
  \begin{align}
    G_1 &\coloneqq \sum_{i=1}^{N} \frac{\bp_i^{2}}{m_i} - N_f kT, \\
    G_j &\coloneqq \frac{p_{\eta_{j-1}}^{2}}{Q_{j-1}} - kT
      \quad (j = 2, \dots, M), \\
    G'_1 &\coloneqq \sum_{c=1}^{n_c} \frac{p_c^{2}}{W_g} - n_c kT, \\
    G'_j &\coloneqq \frac{p_{\xi_{j-1}}^{2}}{Q'_{j-1}} - kT
      \quad (j = 2, \dots, M).
  \end{align}
\end{subequations}
The thermostat and barostat mass parameters are~\cite{Tuckerman2010,Martyna1992,Martyna1994}
\begin{subequations}
  \label{eq:masked-mass}
  \begin{align}
    Q_1 &= N_f kT\tau^{2}, \\
    Q_j &= kT\tau^{2}
      \quad (j = 2, \dots, M), \\
    Q'_1 &= n_c kT\tau^{2}, \\
    Q'_j &= kT\tau^{2}
      \quad (j = 2, \dots, M), \\
    W_g &= \frac{N_f + d}{d}\, kT\tau^{2},
  \end{align}
\end{subequations}
where $\tau$ is the characteristic thermostat/barostat time scale and $Q'_1$ replaces $d^{2}$ with $n_c$ to match the reduced barostat degrees of freedom.
The barostat mass $W_g$ retains its fully anisotropic form rather than being re-derived for the restricted case; this is a pragmatic choice that preserves compatibility with existing implementations.
The masked MTK equations of motion are then
\begin{subequations}
  \label{eq:eom-masked-physical}
  \begin{align}
    \dot{\br}_i &= \frac{\bp_i}{m_i} + \frac{\pg}{W_g} \br_i, \\
    \dot{\bp}_i &= \bF_i
      - \left( \pg + \frac{\Tr[\pg]}{N_f} \bI \right)
        \frac{\bp_i}{W_g}
      - \frac{p_{\eta_1}}{Q_1} \bp_i, \\
    \dot{\bh} &= \frac{\bh \pg}{W_g}, \\
    \dot{p}_c &= \det[\bh] \cdot \be_c^{\top}
      (\mathbf{P}^{\mathrm{int}} - P\bI)\, \be_c
      + \frac{1}{N_f} \sum_{i=1}^{N} \frac{\bp_i^{2}}{m_i}
      - \frac{p_{\xi_1}}{Q'_1}\, p_c
      \quad (c = 1, \dots, n_c),
  \end{align}
\end{subequations}
and the thermostat/barostat NHC chains obey
\begin{subequations}
  \label{eq:eom-masked-chains}
  \begin{align}
    \dot{\eta}_j &= \frac{p_{\eta_j}}{Q_j}
      \quad (j = 1, \dots, M), \\
    \dot{p}_{\eta_j} &= G_j - \frac{p_{\eta_{j+1}}}{Q_{j+1}} p_{\eta_j}
      \quad (j = 1, \dots, M-1), \\
    \dot{p}_{\eta_M} &= G_M, \\
    \dot{\xi}_j &= \frac{p_{\xi_j}}{Q'_j}
      \quad (j = 1, \dots, M), \\
    \dot{p}_{\xi_j} &= G'_j - \frac{p_{\xi_{j+1}}}{Q'_{j+1}} p_{\xi_j}
      \quad (j = 1, \dots, M-1), \\
    \dot{p}_{\xi_M} &= G'_M.
  \end{align}
\end{subequations}
Here $P$ is the target pressure, $\bI$ is the $d \times d$ identity matrix, and only the $n_c$ active cell components fluctuate.
The orthogonality condition [Eq.~\eqref{eq:ortho-assumption}] ensures that $\be_c^{\top} (\mathbf{P}^{\mathrm{int}} - P\bI)\, \be_c$ extracts a single diagonal element of the stress tensor, giving a well-defined scalar pressure along axis $c$.

Although we write as if $\bh$ is fully anisotropic in Eq.~\eqref{eq:eom-masked-physical}, the masked form of $\pg$ ensures that only the $n_c$ active cell lengths fluctuate,
\begin{align}
  \dot{\lambda}_c &= \frac{\lambda_c p_c}{W_g} \quad (c = 1, \dots, n_c).
\end{align}

\subsection{\label{sec:conserved}Conserved energy-like quantity}

The conserved energy-like quantity for the masked MTK barostat is
\begin{align}
  H' &\coloneqq \cH(\br, \bp)
       + \sum_{c=1}^{n_c} \frac{p_c^{2}}{2 W_g}
       + P \det[\bh]
       + \sum_{j=1}^{M}
         \left(
           \frac{p_{\eta_j}^{2}}{2 Q_j}
           + \frac{p_{\xi_j}^{2}}{2 Q'_j}
         \right)
       + N_f kT \eta_1
       + n_c kT \xi_1
       + kT (\eta_c + \xi_c),
  \label{eq:conserved-masked}
\end{align}
where $\eta_c \coloneqq \sum_{j=2}^{M} \eta_j$ and $\xi_c \coloneqq \sum_{j=2}^{M} \xi_j$.
Comparing with the fully anisotropic conserved energy-like quantity [Eq.~\eqref{eq:conserved-aniso}], the only structural differences are:
\begin{enumerate}
  \item The barostat kinetic energy sums over the $n_c$ active components $p_c$ instead of the full $d \times d$ matrix $\pg$.
  \item The coupling to the first barostat chain variable is $n_c kT \xi_1$ instead of $d^{2} kT \xi_1$.
\end{enumerate}

We now verify that $\dot{H}' = 0$ by computing the time derivative and using the equations of motion.

\paragraph*{(I) Physical Hamiltonian, barostat kinetic energy, and $PV$ term.}

Using $\bF_i = -\partial U/\partial \br_i$ and the stress tensor identity [Eq.~\eqref{eq:stress-masked}], the time derivative of the physical Hamiltonian is
\begin{align}
  \dot{\cH} = -\frac{\Tr[\pg]}{N_f W_g} \sum_{i=1}^{N} \frac{\bp_i^{2}}{m_i}
              - \frac{p_{\eta_1}}{Q_1} \sum_{i=1}^{N} \frac{\bp_i^{2}}{m_i}
              - \frac{\det[\bh]}{W_g} \Tr\!\left[ \pg \mathbf{P}^{\mathrm{int}} \right],
  \label{eq:dHdt}
\end{align}
where we use $\pg^\top = \pg$.

The barostat kinetic energy evolves as
\begin{align}
  \frac{d}{dt}\sum_{c=1}^{n_c} \frac{p_c^{2}}{2W_g}
  = \frac{\det[\bh]}{W_g} \Tr\!\left[ \pg \mathbf{P}^{\mathrm{int}} \right]
    - \frac{P\det[\bh]}{W_g} \Tr[\pg]
    + \frac{\Tr[\pg]}{N_f W_g} \sum_{i=1}^{N} \frac{\bp_i^{2}}{m_i}
    - \frac{p_{\xi_1}}{Q'_1} \sum_{c=1}^{n_c} \frac{p_c^{2}}{W_g}.
  \label{eq:dKbaro}
\end{align}
The $PV$ term gives
\begin{align}
  \frac{d}{dt}\bigl(P\det[\bh]\bigr)
  = P\det[\bh]\, \Tr[\bh^{-1}\dot{\bh}]
  = \frac{P\det[\bh]}{W_g} \Tr[\pg].
  \label{eq:dPV}
\end{align}
Summing Eqs.~\eqref{eq:dHdt}--\eqref{eq:dPV}, all stress-tensor and $PV$ terms cancel, leaving
\begin{align}
  \dot{\cH} + \frac{d}{dt}\!\left(\sum_{c=1}^{n_c} \frac{p_c^2}{2W_g}\right) + P\frac{d}{dt}\det[\bh]
  = -\frac{p_{\eta_1}}{Q_1}\sum_{i=1}^{N} \frac{\bp_i^2}{m_i}
    -\frac{p_{\xi_1}}{Q'_1}\sum_{c=1}^{n_c} \frac{p_c^2}{W_g}.
  \label{eq:group1}
\end{align}

\paragraph*{(II) Thermostat chain.}

The time derivative of the thermostat chain terms in $H'$ is
\begin{align}
  \frac{d}{dt}\!\left[
    \sum_{j=1}^{M} \frac{p_{\eta_j}^2}{2Q_j} + N_f kT\eta_1 + kT\eta_c
  \right]
  = +\frac{p_{\eta_1}}{Q_1}\sum_{i=1}^{N} \frac{\bp_i^2}{m_i}.
  \label{eq:group2}
\end{align}

\paragraph*{(III) Barostat chain.}

Similarly, the barostat chain contribution is
\begin{align}
  \frac{d}{dt}\!\left[
    \sum_{j=1}^{M} \frac{p_{\xi_j}^2}{2Q'_j} + n_c kT\xi_1 + kT\xi_c
  \right]
  = +\frac{p_{\xi_1}}{Q'_1}\sum_{c=1}^{n_c} \frac{p_c^2}{W_g}.
  \label{eq:group3}
\end{align}

Summing Eqs.~\eqref{eq:group1}--\eqref{eq:group3}, all terms cancel and $\dot{H}' = 0$.

\subsection{Phase-space compressibility and metric factor}

The extended phase space comprises the variables $\bx = (\{\br_i, \bp_i\}, \{ \lambda_c \}, \{p_c\}, \{\eta_j, p_{\eta_j}\}, \{\xi_j, p_{\xi_j}\})$.
We compute the compressibility $\kappa = \nabla_{\bx}\cdot\dot{\bx}$ by summing the diagonal contributions from each group of variables:
\begin{subequations}
  \begin{align}
    \sum_{i=1}^{N} \sum_{\alpha=1}^{d} \frac{\partial \dot{r}_{i\alpha}}{\partial r_{i\alpha}}
    &= \frac{N\,\Tr[\pg]}{W_g},
    \label{eq:kappa-r}\\
    \sum_{i=1}^{N} \sum_{\alpha=1}^{d} \frac{\partial \dot{p}_{i\alpha}}{\partial p_{i\alpha}}
    &= -\frac{(N+1)\Tr[\pg]}{W_g} - N_f \frac{p_{\eta_1}}{Q_1},
    \label{eq:kappa-p}\\
    \sum_{c=1}^{n_c} \frac{\partial \dot{\lambda}_{c}}{\partial \lambda_{c}}
    &= \frac{\Tr[\pg]}{W_g},
    \label{eq:kappa-lambda}\\
    \sum_{c=1}^{n_c} \frac{\partial \dot{p}_c}{\partial p_c}
    &= -n_c \frac{p_{\xi_1}}{Q'_1},
    \label{eq:kappa-pc}
  \end{align}
\end{subequations}
where Eq.~\eqref{eq:kappa-p} uses $N_f = dN$.
The NHC chains contribute
\begin{align}
  -\sum_{j=2}^{M} \frac{p_{\eta_j}}{Q_j}
  -\sum_{j=2}^{M} \frac{p_{\xi_j}}{Q'_j}.
  \label{eq:kappa-nhc}
\end{align}
Collecting all terms and using $p_{\eta_j}/Q_j = \dot{\eta}_j$, $p_{\xi_j}/Q'_j = \dot{\xi}_j$:
\begin{align}
  \kappa = -\frac{d}{dt}\!\left(
    N_f\eta_1 + n_c\xi_1 + \eta_c + \xi_c
  \right).
  \label{eq:kappa-masked}
\end{align}
Since $\kappa = \dot{w}$, the metric factor is
\begin{align}
  e^{-w}
  = \exp\left(
      N_f\eta_1 + n_c\xi_1 + \eta_c + \xi_c
    \right).
  \label{eq:metric-masked}
\end{align}

Geometrically, the masked barostat momentum $\pg$ [Eq.~\eqref{eq:pg-masked}] lives in the $n_c$-dimensional linear subspace spanned by $\{\be_c \be_c^{\top}\}$, so the first barostat-chain entropic term ($n_c \xi_1$) and the metric factor [Eq.~\eqref{eq:metric-masked}] each count $n_c$ degrees of freedom rather than the $d^{2}$ of the fully anisotropic case.

\subsection{\label{sec:ensemble}Isothermal--isobaric ensemble}

The microcanonical partition function is
\begin{align}
  \Omega(E')
    &= \int d\bx\, e^{-w} \delta(H'(\bx) - E') \\
    &= \frac{1}{n_c kT}
      \left( \prod_{c=1}^{n_c} \sqrt{2 \pi k T W_g} \right)
      \left( \prod_{j=1}^{M} \sqrt{2 \pi k T Q_j} \right)
      \left( \prod_{j=1}^{M} \sqrt{2 \pi k T Q'_j} \right) \nonumber \\
      &\quad
      \int d\br d\bp d\lambda^{n_c} d\eta^{M} d\xi^{M-1}
        \exp\left( -\frac{1}{kT} \left( \cH(\br, \bp) + P \det[\bh] - E' \right) \right).
\end{align}
In passing from the first to the second line, the barostat momenta $p_c$ ($c = 1, \dots, n_c$), the thermostat momenta $p_{\eta_j}$, $p_{\xi_j}$, and the thermostat positions $\eta_j$ ($j \geq 2$), $\xi_j$ ($j \geq 2$) are integrated out as Gaussian integrals, producing the square-root prefactors.
The $\eta_1$ integral enforces the energy-shell constraint via $\delta(H' - E')$, and because the conserved energy-like quantity $H'$ [Eq.~\eqref{eq:conserved-masked}] contains the term $n_c kT \xi_1$, integrating $\xi_1$ over the delta function yields the prefactor $1/(n_c kT)$---directly analogous to the $1/(d^{2} kT)$ factor in the fully anisotropic case.
Only the $n_c$ active cell lengths $\lambda_c$ appear as dynamical variables in $H'$; the inactive cell lengths $\lambda_k$ ($k = n_c + 1, \dots, d$) satisfy $\dot{\lambda}_k = 0$ and are absent from the conserved energy-like quantity, so they contribute no integration measure.
These inactive axes are externally fixed parameters determined by the initial condition, not sampled variables; they enter $\det[\bh]$ only as constant multiplicative factors.

Because the prefactors are independent of the real dynamical variables $\br$, $\bp$, and $\lambda_c$, the microcanonical partition function is proportional to the isothermal--isobaric partition function,
\begin{align}
  \Omega(E')
    &\propto
      \int d\br d\bp d\lambda^{n_c}
        \exp\left( -\frac{1}{kT} \left( \cH(\br, \bp) + P \det[\bh] \right) \right)
      =: \Delta(N, P, T).
\end{align}
Thus, the masked MTK equations sample the isothermal--isobaric ensemble $(N, P, T)$ restricted to the submanifold in which the $d - n_c$ inactive cell axes are held fixed.

\subsection{Isothermal--isobaric ensemble without external forces}

When there is no external field $\sum_{i=1}^{N} \bF_i = \mathbf{0}$, the total momentum $\bP = \sum_{i=1}^{N} \bp_i$ obeys
\begin{align}
  \dot{\bP}
    &= -\frac{1}{W_g} \pg \bP
      -\left(
        \frac{\Tr[\pg]}{W_g} + \frac{p_{\eta_1}}{Q_1}
      \right) \bP \nonumber \\
    &= - \frac{d}{dt} \left(
      \log [\bh] + \frac{\log \det [\bh]}{N_f} \bI + \eta_1 \bI
    \right) \bP.
\end{align}
Thus, the additional momentum-like quantity
\begin{align}
  \bK(\bx) &\coloneqq \bh \bP (\det [\bh])^{1/N_f} e^{\eta_1}
\end{align}
is also conserved.
Because $\bK$ is a constant of motion, the trajectory is confined to a fixed-$\bK$ submanifold of the extended phase space.
To ensure that the dynamics still samples the correct isothermal--isobaric ensemble under this additional constraint, we must verify that the microcanonical partition function $\Omega(E', \bK')$ remains proportional to $\Delta(N, P, T)$.
We write the values as $\bK' = \bK(\bx)$.

When $\bK' \neq \bm{0}$, the microcanonical partition function is
\begin{align}
  \Omega(E', \bK')
    &= \int d\bx\, e^{-w} \delta(H'(\bx) - E') \delta^{(d)}( [\bh \bP] (\det [\bh])^{1/N_f} e^{\eta_1} - \bK' ) \\
    &= \frac{1}{n_c kT}
      \left( \prod_{c=1}^{n_c} \sqrt{2 \pi k T W_g} \right)
      \left( \prod_{j=1}^{M} \sqrt{2 \pi k T Q_j} \right)
      \left( \prod_{j=1}^{M} \sqrt{2 \pi k T Q'_j} \right) \nonumber \\
      &\quad
      \int d\br d\bp d\lambda^{n_c} d\eta^{M} d\xi^{M-1}
        \exp\left( -\frac{1}{kT} \left( \cH(\br, \bp) + P \det[\bh] - E' \right) \right)
        \delta^{(d)}\left( \bh \bP (\det [\bh])^{1/N_f} e^{\eta_1} - \bK' \right).
\end{align}
The first-to-second-line step performs the same Gaussian integrations as in the $\bK'$-unconstrained case: the barostat momenta $p_c$, thermostat momenta $p_{\eta_j}$, $p_{\xi_j}$, and the $\xi_1$ and $\eta_1$ integrals produce the square-root prefactors and $1/(n_c kT)$, while the delta function $\delta^{(d)}(\bK - \bK')$ passes through unchanged because none of the integrated-out variables appear in $\bK$.
In the third line, the remaining thermostat positions $\eta_j$ ($j \geq 2$) and $\xi_j$ ($j \geq 2$) are absorbed into the proportionality constant, leaving the integration over $d\br\, d\bp\, d\lambda^{n_c}\, d\eta_1$ with the delta-function constraint on $\bK'$.

The $\eta_1$ integral is then evaluated against $\delta^{(d)}(\bK(\bx) - \bK')$; since $\bK$ depends on $\eta_1$ only through the overall factor $e^{\eta_1}$, the substitution $u_\alpha = K'_\alpha e^{-\eta_1}$ produces the Jacobian $1/\prod_\alpha |K'_\alpha|$ and eliminates $\eta_1$,
\begin{align}
  \Omega(E', \bK')
    &\propto
      \int d\br d\bp d\lambda^{n_c} d\eta_1
        \exp\left( -\frac{1}{kT} \left( \cH(\br, \bp) + P \det[\bh] \right) \right)
        \delta^{(d)}\left( \bh \bP (\det [\bh])^{1/N_f} e^{\eta_1} - \bK' \right)
    \nonumber \\
    &=
      \frac{1}{\prod_{\alpha} |K'_{\alpha}|}
      \int d\br d\bp d\lambda^{n_c}
        \exp\left( -\frac{1}{kT} \left( \cH(\br, \bp) + P \det[\bh] \right) \right) \\
    &\propto \Delta(N, P, T). \nonumber
\end{align}
This prefactor is independent of the remaining integration variables, so it factors out and the surviving integral is $\Delta(N, P, T)$.

When $\bK(\bx) = \bm{0}$, the total momentum $\bP$ is always zero and the microcanonical partition function is
\begin{align}
  \Omega(E', \bK') &\propto \Omega(E') \propto \Delta(N, P, T).
\end{align}
Thus, in either case, the masked MTK equations sample the isothermal--isobaric ensemble $(N, P, T)$ restricted to the submanifold in which the inactive cell axes are held fixed.

\section{Summary}
\label{sec:summary}

We have derived the conserved energy-like quantity for MTK barostats with a restricted subset of $n_c$ active cell degrees of freedom [Eq.~\eqref{eq:conserved-masked}] and verified that it is exactly conserved (Sec.~\ref{sec:conserved}).
Using the generalized Liouville framework, we computed the phase-space compressibility and metric factor for the restricted system [Eqs.~\eqref{eq:kappa-masked}--\eqref{eq:metric-masked}] and showed that the resulting dynamics samples the isothermal--isobaric ensemble restricted to the submanifold of cell shapes with inactive axes held fixed (Sec.~\ref{sec:ensemble}).
The key structural difference from the fully anisotropic case is that $d^{2}$ is replaced by $n_c$ in every term that counts barostat degrees of freedom---the barostat NHC driving force, the NHC mass parameter, the conserved energy-like quantity, and the metric factor.
This result provides a compact reference for verifying energy conservation and ensemble correctness in restricted-axis isobaric simulations.
The orthogonality assumption on the active axes [Eq.~\eqref{eq:ortho-assumption}] limits the method to cell shapes where each active axis is perpendicular to all others; non-orthogonal active subspaces such as hexagonal in-plane pairs require an orthohexagonal redefinition, while fully triclinic cells are excluded altogether.
We also provide a complete Liouville-operator-based integration scheme for the restricted MTK barostat (Appendix~\ref{app:masked-integrator}), constructed from the same operator-splitting approach used for the isotropic and fully anisotropic cases summarized for completeness in Appendices~\ref{app:iso}--\ref{app:aniso}.

\section{Acknowledgments}

The author acknowledges the use of the AI-based writing tool ChatGPT (OpenAI, version 5.4) and Claude (Anthropic, version 4.6) for language polishing.
The author takes full responsibility for the content of this manuscript.

\appendix

\section{Isothermal integrator with Nos\'e--Hoover chains}

\label{app:nhc}

\subsection{Equations of motion, conserved energy-like quantity, and conserved momentum-like quantity}

Section~\ref{sec:masked} already includes the thermostat and barostat chains used in the masked MTK dynamics.
This appendix isolates the standard Nos\'e--Hoover chain (NHC) thermostat, which is reused unchanged in both the masked and fully anisotropic formulations~\cite{Martyna1992,Tuckerman2010}.
\begin{subequations}
  \begin{align}
    \dot{\br}_i &= \frac{\bp_i}{m_i} \\
    \dot{\bp}_i &= \bF_i - \frac{p_{\eta_1}}{Q_1} \bp_i,
  \end{align}
\end{subequations}
together with the thermostat chain equations
\begin{subequations}
  \label{eq:nhc-eom}
  \begin{align}
    \dot{\eta}_j &= \frac{p_{\eta_j}}{Q_j}
      \quad (j = 1, \dots, M) \\
    \dot{p}_{\eta_1} &= \sum_{i=1}^{N} \frac{\bp_i^{2}}{m_i} - N_f kT
      - \frac{p_{\eta_2}}{Q_2} p_{\eta_1} \\
    \dot{p}_{\eta_j} &= \frac{p_{\eta_{j-1}}^{2}}{Q_{j-1}} - kT
      - \frac{p_{\eta_{j+1}}}{Q_{j+1}} p_{\eta_j}
      \quad (j = 2, \dots, M-1) \\
    \dot{p}_{\eta_M} &= \frac{p_{\eta_{M-1}}^{2}}{Q_{M-1}} - kT,
  \end{align}
\end{subequations}
where $M$ is the chain length.
The thermostat masses $Q_j$ and time scale $\tau$ are as defined in Eq.~\eqref{eq:masked-mass}.

The NHC conserved energy-like quantity is
\begin{align}
  H' \coloneqq \cH(\br, \bp)
    + \sum_{j=1}^{M} \frac{p_{\eta_j}^{2}}{2 Q_j}
    + N_f kT \eta_1
    + kT \eta_c,
  \label{eq:nhc-conserved}
\end{align}
where $\eta_c$ is as defined after Eq.~\eqref{eq:conserved-masked}.

In the absence of external forces, the total momentum $\bP = \sum_{i=1}^{N} \bp_i$ has the conserved momentum-like quantity
\begin{align}
  \bK(\bx) \coloneqq \bP e^{\eta_1}.
\end{align}

The phase-space compressibility and metric factor are
\begin{align}
  \kappa &= -\frac{d}{dt} \left( N_f \eta_1 + \eta_c \right) \\
  e^{-w} &= \exp\left( N_f \eta_1 + \eta_c \right).
\end{align}

\subsection{Integration scheme based on Liouville operator factorization}

The Liouville operator for the NHC thermostat is decomposed as
\begin{subequations}
  \label{eq:nhc-liouville}
  \begin{align}
    i\mathcal{L} &= i\mathcal{L}_{\mathrm{NHC}} + i\mathcal{L}_1 + i\mathcal{L}_2 \\
    i\mathcal{L}_1 &\coloneqq \sum_{i=1}^{N} \frac{\bp_i}{m_i}
      \cdot \frac{\partial}{\partial \br_i} \\
    i\mathcal{L}_2 &\coloneqq \sum_{i=1}^{N} \bF_i
      \cdot \frac{\partial}{\partial \bp_i} \\
    i\mathcal{L}_{\mathrm{NHC}} &\coloneqq
      - \sum_{i=1}^{N} \frac{p_{\eta_1}}{Q_1} \bp_i
        \cdot \frac{\partial}{\partial \bp_i}
      + \sum_{j=1}^{M} \frac{p_{\eta_j}}{Q_j}
        \frac{\partial}{\partial \eta_j}
      + \sum_{j=1}^{M-1}
        \left( G_j - p_{\eta_j} \frac{p_{\eta_{j+1}}}{Q_{j+1}} \right)
        \frac{\partial}{\partial p_{\eta_j}}
      + G_M \frac{\partial}{\partial p_{\eta_M}},
  \end{align}
\end{subequations}
with driving forces $G_j$ as defined in Eq.~\eqref{eq:masked-driving}.

The second-order Trotter factorization gives the propagator
\begin{align}
  e^{i\mathcal{L} \Delta t}
  = e^{i\mathcal{L}_{\mathrm{NHC}} \frac{\Delta t}{2}}\,
    e^{i\mathcal{L}_2 \frac{\Delta t}{2}}\,
    e^{i\mathcal{L}_1 \Delta t}\,
    e^{i\mathcal{L}_2 \frac{\Delta t}{2}}\,
    e^{i\mathcal{L}_{\mathrm{NHC}} \frac{\Delta t}{2}}
    + \cO(\Delta t^{3}).
  \label{eq:nhc-trotter}
\end{align}
The actions of $e^{i\mathcal{L}_1 \Delta t}$ and $e^{i\mathcal{L}_2 \frac{\Delta t}{2}}$ are
\begin{subequations}
  \begin{align}
    e^{i\mathcal{L}_1 \Delta t}
      \begin{pmatrix} \br_i \\ \bp_i \\ \eta_j \\ p_{\eta_j} \end{pmatrix}
    &=
      \begin{pmatrix}
        \br_i + \frac{\bp_i}{m_i} \Delta t \\
        \bp_i \\ \eta_j \\ p_{\eta_j}
      \end{pmatrix}
    \label{eq:nhc-action-L1} \\
    e^{i\mathcal{L}_2 \frac{\Delta t}{2}}
      \begin{pmatrix} \br_i \\ \bp_i \\ \eta_j \\ p_{\eta_j} \end{pmatrix}
    &=
      \begin{pmatrix}
        \br_i \\
        \bp_i + \bF_i \frac{\Delta t}{2} \\
        \eta_j \\ p_{\eta_j}
      \end{pmatrix}.
    \label{eq:nhc-action-L2}
  \end{align}
\end{subequations}

The NHC propagator is further decomposed via a Suzuki--Yoshida factorization~\cite{Yoshida1990}.
Denoting $n$ sub-steps per half time step,
\begin{align}
  e^{i\mathcal{L}_{\mathrm{NHC}} \frac{\Delta t}{2}}
  &= \left( e^{i\mathcal{L}_{\mathrm{NHC}} \frac{\Delta t}{2n}} \right)^{n}
  \nonumber \\
  e^{i\mathcal{L}_{\mathrm{NHC}} \frac{\Delta t}{2n}}
  &= S_4^{\mathrm{NHC}}\!\left( \frac{\Delta t}{2n} \right)
     + \cO\!\left( \left(\frac{\Delta t}{n}\right)^{5} \right)
  \nonumber \\
  S_4^{\mathrm{NHC}}\!\left( \frac{\Delta t}{2n} \right)
  &\coloneqq \prod_{\alpha=1}^{3}
    S_2^{\mathrm{NHC}}
      \!\left( x_\alpha \frac{\Delta t}{2n} \right),
  \label{eq:nhc-sy}
\end{align}
with the Suzuki--Yoshida weights $x_\alpha$.
Writing $\delta_\alpha \coloneqq x_\alpha \Delta t / (2n)$, the second-order kernel is
\begin{align}
  S_2^{\mathrm{NHC}}(\delta_\alpha)
  &\coloneqq
    \exp\!\left( \frac{\delta_\alpha}{2} G_M
      \frac{\partial}{\partial p_{\eta_M}} \right)
  \nonumber \\
  &\quad \times
    \prod_{j=M-1}^{1}
    \left[
      \exp\!\left( -\frac{\delta_\alpha}{4}
        \frac{p_{\eta_{j+1}}}{Q_{j+1}} p_{\eta_j}
        \frac{\partial}{\partial p_{\eta_j}} \right)
      \exp\!\left( \frac{\delta_\alpha}{2} G_j
        \frac{\partial}{\partial p_{\eta_j}} \right)
      \exp\!\left( -\frac{\delta_\alpha}{4}
        \frac{p_{\eta_{j+1}}}{Q_{j+1}} p_{\eta_j}
        \frac{\partial}{\partial p_{\eta_j}} \right)
    \right]
  \nonumber \\
  &\quad \times
    \prod_{i=1}^{N}
    \exp\!\left( -\delta_\alpha
      \frac{p_{\eta_1}}{Q_1} \bp_i
      \cdot \frac{\partial}{\partial \bp_i} \right)
  \nonumber \\
  &\quad \times
    \prod_{j=1}^{M}
    \exp\!\left( \delta_\alpha
      \frac{p_{\eta_j}}{Q_j}
      \frac{\partial}{\partial \eta_j} \right)
  \nonumber \\
  &\quad \times
    \prod_{j=1}^{M-1}
    \left[
      \exp\!\left( -\frac{\delta_\alpha}{4}
        \frac{p_{\eta_{j+1}}}{Q_{j+1}} p_{\eta_j}
        \frac{\partial}{\partial p_{\eta_j}} \right)
      \exp\!\left( \frac{\delta_\alpha}{2} G_j
        \frac{\partial}{\partial p_{\eta_j}} \right)
      \exp\!\left( -\frac{\delta_\alpha}{4}
        \frac{p_{\eta_{j+1}}}{Q_{j+1}} p_{\eta_j}
        \frac{\partial}{\partial p_{\eta_j}} \right)
    \right]
  \nonumber \\
  &\quad \times
    \exp\!\left( \frac{\delta_\alpha}{2} G_M
      \frac{\partial}{\partial p_{\eta_M}} \right).
  \label{eq:nhc-s2}
\end{align}
Here, we use the identity
\begin{align}
  \exp\!\left( c\, x \frac{\partial}{\partial x} \right) f(x) = f(x\, e^{c})
  \label{eq:exp-scaling}
\end{align}
to evaluate the actions of each exponential operator.

\section{Isothermal--isobaric integrator with isotropic MTK barostat}
\label{app:iso}

\subsection{Equations of motion, conserved energy-like quantity, and conserved momentum-like quantity}

The MTK equations for isotropic volume fluctuations are~\cite{Martyna1994,Tuckerman2010}
\begin{subequations}
  \label{eq:eom-iso}
  \begin{align}
    \dot{\br}_i &= \frac{\bp_i}{m_i} + \frac{p_\epsilon}{W} \br_i \\
    \dot{\bp}_i &= \bF_i
      - \left(1 + \frac{d}{N_f}\right) \frac{p_\epsilon}{W} \bp_i
      - \frac{p_{\eta_1}}{Q_1} \bp_i \\
    \dot{V} &= \frac{d V}{W} p_\epsilon \\
    \dot{p}_\epsilon &= d V (P^{\mathrm{int}} - P)
      + \frac{d}{N_f} \sum_{i=1}^{N} \frac{\bp_i^{2}}{m_i}
      - \frac{p_{\xi_1}}{Q'_1} p_\epsilon,
  \end{align}
\end{subequations}
together with the same thermostat chain equations as in Eq.~\eqref{eq:nhc-eom} and barostat chain equations of the same form as Eq.~\eqref{eq:eom-masked-chains}, with driving forces
\begin{subequations}
  \label{eq:driving-iso}
  \begin{align}
    G'_1 &\coloneqq \frac{p_\epsilon^{2}}{W} - kT \\
    G'_j &\coloneqq \frac{p_{\xi_{j-1}}^{2}}{Q'_{j-1}} - kT
      \quad (j = 2, \dots, M).
  \end{align}
\end{subequations}
Here $P^{\mathrm{int}}$ denotes the isotropic part of the internal stress tensor, $P^{\mathrm{int}} \coloneqq \frac{1}{d} \Tr[\mathbf{P}^{\mathrm{int}}]$.
The thermostat driving forces $G_1$, $G_j$ and thermostat masses $Q_1$, $Q_j$ are the same as in Appendix~\ref{app:nhc}.

Reference~\onlinecite{Martyna1994} suggests $W = (N_f + d)\, kT\tau^{2}$ and $Q'_1 = kT\tau^{2}$; the remaining barostat chain masses $Q'_j$ ($j \geq 2$) and time scale $\tau$ are as in Eq.~\eqref{eq:masked-mass}.

The conserved energy-like quantity of the isotropic MTK equations is
\begin{align}
  H' &\coloneqq \cH(\br, \bp)
    + \frac{p_\epsilon^{2}}{2 W} + P V
    + \sum_{j=1}^{M}
      \left(
        \frac{p_{\eta_j}^{2}}{2 Q_j}
        + \frac{p_{\xi_j}^{2}}{2 Q'_j}
      \right)
    + N_f kT \eta_1 + kT \xi_1 + kT \eta_c + kT \xi_c,
  \label{eq:conserved-iso}
\end{align}
where $\eta_c$ and $\xi_c$ are as defined after Eq.~\eqref{eq:conserved-masked}.

In the absence of external forces ($\sum_{i=1}^{N} \bF_i = 0$), the total momentum $\bP = \sum_{i=1}^{N} \bp_i$ has the conserved quantity
\begin{align}
  \bK \coloneqq \bP \exp\!\left(
    \left(1 + \frac{d}{N_f}\right) \epsilon + \eta_1
  \right),
  \label{eq:conserved-momentum-iso}
\end{align}
where $\epsilon \coloneqq \frac{1}{d} \ln \frac{V}{V_0}$ with some reference volume $V_0$.

The phase-space compressibility and metric factor are
\begin{align}
  \kappa &= -\frac{d}{dt}\!\left(
    N_f \eta_1 + \xi_1 + \eta_c + \xi_c
  \right) \\
  e^{-w} &= \exp\left(
    N_f \eta_1 + \xi_1 + \eta_c + \xi_c
  \right).
\end{align}

\subsection{Integration scheme based on Liouville operator factorization}

The Liouville operator for the isotropic MTK equations is decomposed as
\begin{align}
  i\mathcal{L}
    = i\mathcal{L}_1 + i\mathcal{L}_2 + i\mathcal{L}_{\epsilon,1} + i\mathcal{L}_{\epsilon,2}
    + i\mathcal{L}_{\mathrm{NHC\text{-}baro}} + i\mathcal{L}_{\mathrm{NHC\text{-}thermo}},
  \label{eq:liouville-decomp-iso}
\end{align}
where
\begin{subequations}
  \label{eq:liouville-parts-iso}
  \begin{align}
    i\mathcal{L}_1 &\coloneqq \sum_{i=1}^{N}
      \left( \frac{\bp_i}{m_i} + \frac{p_\epsilon}{W} \br_i \right)
      \cdot \frac{\partial}{\partial \br_i} \\
    i\mathcal{L}_2 &\coloneqq \sum_{i=1}^{N}
      \left(
        \bF_i
        - \left(1 + \frac{d}{N_f}\right) \frac{p_\epsilon}{W} \bp_i
      \right)
      \cdot \frac{\partial}{\partial \bp_i} \\
    i\mathcal{L}_{\epsilon,1} &\coloneqq \frac{p_\epsilon}{W}
      \frac{\partial}{\partial \epsilon} \\
    i\mathcal{L}_{\epsilon,2} &\coloneqq G_\epsilon
      \frac{\partial}{\partial p_\epsilon}
  \end{align}
\end{subequations}
with the barostat driving force
\begin{align}
  G_\epsilon
    \coloneqq d V (P^{\mathrm{int}} - P) + \frac{d}{N_f} \sum_{i=1}^{N} \frac{\bp_i^{2}}{m_i}.
  \label{eq:Gepsilon}
\end{align}
The NHC-thermostat and NHC-barostat parts $i\mathcal{L}_{\mathrm{NHC\text{-}thermo}}$ and $i\mathcal{L}_{\mathrm{NHC\text{-}baro}}$ have the same structure as in Eq.~\eqref{eq:nhc-liouville}, acting on the thermostat chain $(\eta_j, p_{\eta_j})$ and barostat chain $(\xi_j, p_{\xi_j})$ variables, respectively.
For $i\mathcal{L}_{\mathrm{NHC\text{-}baro}}$, the particle momentum $\bp_i$ in the drag term is replaced by the scalar barostat momentum $p_\epsilon$:
\begin{align}
  i\mathcal{L}_{\mathrm{NHC\text{-}baro}}
    &\coloneqq - \frac{p_{\xi_1}}{Q'_1} p_\epsilon \frac{\partial}{\partial p_\epsilon}
      + \sum_{j=1}^{M} \frac{p_{\xi_j}}{Q'_j} \frac{\partial}{\partial \xi_j}
      + \sum_{j=1}^{M-1}
        \left( G'_j - p_{\xi_j} \frac{p_{\xi_{j+1}}}{Q'_{j+1}} \right)
        \frac{\partial}{\partial p_{\xi_j}}
      + G'_M \frac{\partial}{\partial p_{\xi_M}}.
  \label{eq:nhc-baro-iso}
\end{align}

The second-order Trotter factorization gives
\begin{align}
  e^{i\mathcal{L} \Delta t}
  &= e^{i\mathcal{L}_{\mathrm{NHC\text{-}baro}} \frac{\Delta t}{2}}\,
     e^{i\mathcal{L}_{\mathrm{NHC\text{-}thermo}} \frac{\Delta t}{2}}
     e^{i\mathcal{L}_{\epsilon,2} \frac{\Delta t}{2}}\,
     e^{i\mathcal{L}_2 \frac{\Delta t}{2}}
     e^{i\mathcal{L}_{\epsilon,1} \Delta t}\,
     e^{i\mathcal{L}_1 \Delta t}
     e^{i\mathcal{L}_2 \frac{\Delta t}{2}}\,
     e^{i\mathcal{L}_{\epsilon,2} \frac{\Delta t}{2}}
     e^{i\mathcal{L}_{\mathrm{NHC\text{-}thermo}} \frac{\Delta t}{2}}\,
     e^{i\mathcal{L}_{\mathrm{NHC\text{-}baro}} \frac{\Delta t}{2}}
     + \cO(\Delta t^{3}).
  \label{eq:trotter-iso}
\end{align}

The actions of $e^{i\mathcal{L}_1 \Delta t}$ and $e^{i\mathcal{L}_2 \frac{\Delta t}{2}}$ are
\begin{subequations}
  \begin{align}
    e^{i\mathcal{L}_1 \Delta t} \br_i
    &= \br_i\, e^{\frac{p_\epsilon \Delta t}{W}}
      + \Delta t\, \frac{\bp_i}{m_i}\,
        \exprel\!\left( \frac{p_\epsilon \Delta t}{W} \right),
    \label{eq:action-L1-iso} \\
    e^{i\mathcal{L}_2 \frac{\Delta t}{2}} \bp_i
    &= \bp_i\, e^{-\frac{\kappa \Delta t}{2 W}}
      + \frac{\Delta t}{2}\, \bF_i\,
        \exprel\!\left( -\frac{\kappa \Delta t}{2 W} \right),
    \label{eq:action-L2-iso}
  \end{align}
\end{subequations}
where $\kappa \coloneqq (1 + d/N_f)\, p_\epsilon$ and $\exprel(x) \coloneqq (e^{x} - 1)/x$.

\section{Isothermal--isobaric integrator with fully anisotropic MTK barostat}
\label{app:aniso}

\subsection{Equations of motion, conserved energy-like quantity, and conserved momentum-like quantity}

Relative to the masked system, the particle and cell equations are unchanged; only the barostat momentum is promoted from the diagonal masked form [Eq.~\eqref{eq:pg-masked}] to a full matrix variable.
The MTK equations for fully anisotropic cell fluctuations are~\cite{Martyna1994,Tuckerman2006}
\begin{subequations}
  \label{eq:eom-aniso}
  \begin{align}
    \dot{\br}_i &= \frac{\bp_i}{m_i} + \frac{\pg}{W_g} \br_i \\
    \dot{\bp}_i &= \bF_i
      - \left( \pg + \frac{\Tr[\pg]}{N_f} \bI \right)
        \frac{\bp_i}{W_g}
      - \frac{p_{\eta_1}}{Q_1} \bp_i \\
    \dot{\bh} &= \frac{\bh \pg}{W_g} \\
    \dot{\pg} &= \det[\bh]
      \left( \mathbf{P}^{\mathrm{int}} - P \bI \right)
      + \frac{1}{N_f} \sum_{i=1}^{N} \frac{\bp_i^{2}}{m_i} \bI
      - \frac{p_{\xi_1}}{Q'_1} \pg,
  \end{align}
\end{subequations}
together with the same thermostat chain equations as in Eq.~\eqref{eq:nhc-eom} and the barostat chain equations obtained by replacing $G'_1$ in Eq.~\eqref{eq:eom-masked-chains} with the fully anisotropic driving force below.
Here $P$ is the target pressure and $\bI$ is the $d \times d$ identity matrix.

The thermostat driving forces $G_1$, $G_j$ and thermostat masses $Q_1$, $Q_j$ are the same as in Sec.~\ref{sec:eom} and Appendix~\ref{app:nhc}.
The anisotropic barostat differs only in the first driving force, $G'_1 \coloneqq \Tr[\pg^{\top} \pg] / W_g - d^{2} kT$; the remaining forces $G'_j$ ($j \geq 2$) are as in Eq.~\eqref{eq:masked-driving}.
The cell mass $W_g$ and remaining chain masses $Q'_j$ ($j \geq 2$) are as in Eq.~\eqref{eq:masked-mass}; the first barostat chain mass is $Q'_1 = d^{2}\, kT\tau^{2}$, where $d^{2}$ counts the independent components of $\pg$.

The conserved energy-like quantity of the fully anisotropic MTK equations is~\cite{Martyna1994,Tuckerman2006}
\begin{align}
  H' &\coloneqq \cH(\br, \bp)
       + \frac{\Tr[\pg^{\top} \pg]}{2 W_g}
       + P \det[\bh]
       + \sum_{j=1}^{M}
         \left(
           \frac{p_{\eta_j}^{2}}{2 Q_j}
           + \frac{p_{\xi_j}^{2}}{2 Q'_j}
         \right)
       + N_f kT \eta_1
       + d^{2} kT \xi_1
       + kT \eta_c + kT \xi_c.
  \label{eq:conserved-aniso}
\end{align}
The term $d^{2} kT \xi_1$ reflects that all $d^{2}$ components of $\pg$ are coupled to the first barostat chain variable.

In the absence of external forces ($\sum_{i=1}^{N} \bF_i = \mathbf{0}$), the total momentum $\bP = \sum_{i=1}^{N} \bp_i$ has the conserved quantity
\begin{align}
  \bK \coloneqq \bh\, \bP\, (\det[\bh])^{1/N_f}\, e^{\eta_1}.
  \label{eq:conserved-momentum-aniso}
\end{align}

The phase-space compressibility and metric factor are
\begin{align}
  \kappa &= -\frac{d}{dt}\!\left(
    N_f \eta_1 + d^{2} \xi_1 + \eta_c + \xi_c
  \right) \\
  e^{-w} &= \exp\left(
    N_f \eta_1 + d^{2} \xi_1 + \eta_c + \xi_c
  \right).
\end{align}

\subsection{Integration scheme based on Liouville operator factorization}

We outline the Liouville-operator-based integration scheme for the anisotropic and masked MTK equations, following Refs.~\cite{Tuckerman2006,Tuckerman2010}.

The Liouville operator for the full anisotropic MTK equations is decomposed as~\cite{Tuckerman2006}
\begin{align}
  i\mathcal{L}
    = i\mathcal{L}_1 + i\mathcal{L}_2 + i\mathcal{L}_{g,1} + i\mathcal{L}_{g,2}
       + i\mathcal{L}_{\mathrm{NHC\text{-}baro}} + i\mathcal{L}_{\mathrm{NHC\text{-}thermo}},
  \label{eq:liouville-decomp}
\end{align}
where
\begin{subequations}
  \label{eq:liouville-parts}
  \begin{align}
    i\mathcal{L}_1 &\coloneqq \sum_{i=1}^{N}
      \left( \frac{\bp_i}{m_i} + \frac{\pg}{W_g} \br_i \right)
      \cdot \frac{\partial}{\partial \br_i} \\
    i\mathcal{L}_2 &\coloneqq \sum_{i=1}^{N}
      \left(
        \bF_i
        - \left( \pg + \frac{\Tr[\pg]}{N_f} \bI \right)
          \frac{\bp_i}{W_g}
      \right)
      \cdot \frac{\partial}{\partial \bp_i} \\
    i\mathcal{L}_{g,1} &\coloneqq \frac{\bh \pg}{W_g} : \frac{\partial}{\partial \bh} \\
    i\mathcal{L}_{g,2} &\coloneqq \Gg : \frac{\partial}{\partial \pg}.
  \end{align}
\end{subequations}
Here $\mathbf{A} : \mathbf{B}$ denotes the double contraction $\Tr[\mathbf{A}^{\top} \mathbf{B}]$ between matrices $\mathbf{A}$ and $\mathbf{B}$.
The barostat driving force matrix is
\begin{align}
  \Gg \coloneqq \det[\bh]
    \left( \mathbf{P}^{\mathrm{int}} - P\bI \right)
    + \frac{1}{N_f} \sum_{i=1}^{N} \frac{\bp_i^{2}}{m_i}\, \bI.
  \label{eq:Gg}
\end{align}
The second-order Trotter factorization gives the propagator~\cite{Tuckerman2006} (a fourth-order scheme can be constructed via the Suzuki--Yoshida decomposition~\cite{Yoshida1990})
\begin{align}
  e^{i\mathcal{L}\Delta t}
  &= e^{i\mathcal{L}_{\mathrm{NHC\text{-}baro}} \frac{\Delta t}{2}}\,
     e^{i\mathcal{L}_{\mathrm{NHC\text{-}thermo}} \frac{\Delta t}{2}}
     e^{i\mathcal{L}_{g,2} \frac{\Delta t}{2}}\,
     e^{i\mathcal{L}_2 \frac{\Delta t}{2}}
     e^{i\mathcal{L}_{g,1} \Delta t}\,
     e^{i\mathcal{L}_1 \Delta t}
     e^{i\mathcal{L}_2 \frac{\Delta t}{2}}\,
     e^{i\mathcal{L}_{g,2} \frac{\Delta t}{2}}
     e^{i\mathcal{L}_{\mathrm{NHC\text{-}thermo}} \frac{\Delta t}{2}}\,
     e^{i\mathcal{L}_{\mathrm{NHC\text{-}baro}} \frac{\Delta t}{2}}
     + \cO(\Delta t^{3}).
  \label{eq:trotter}
\end{align}

Since $\pg$ is a real symmetric matrix, it admits an eigendecomposition
\begin{align}
  \pg = \sum_{\mu=1}^{d} \rho_\mu \bu_\mu \bu_\mu^{\top},
  \qquad
  \bU \coloneqq (\bu_1, \dots, \bu_d),
  \label{eq:pg-eigen}
\end{align}
with $\rho_\mu \in \mathbb{R}$ and $\bu_\mu^{\top} \bu_\nu = \delta_{\mu\nu}$.
Introducing scaled coordinates $\bx_i \coloneqq \bU^{\top} \br_i$ and $\by_i \coloneqq \bU^{\top} \bp_i$, the actions of $e^{i\mathcal{L}_1 \Delta t}$ and $e^{i\mathcal{L}_2 \frac{\Delta t}{2}}$ decouple along each eigenvector direction~\cite{Tuckerman2006}:
\begin{subequations}
  \begin{align}
    e^{i\mathcal{L}_1 \Delta t} \br_i
      &= \bU
        \left(
          x_{i\alpha}\, e^{\frac{\rho_\alpha \Delta t}{W_g}}
          + \Delta t\, \frac{[\bU^{\top} \bp_i]_\alpha}{m_i}\,
            \exprel\!\left( \frac{\rho_\alpha \Delta t}{W_g} \right)
        \right)_{\!\alpha=1,\dots,d}
      \label{eq:action-L1} \\
    e^{i\mathcal{L}_2 \frac{\Delta t}{2}} \bp_i
      &= \bU
        \left(
          y_{i\alpha}\, e^{-\frac{\kappa_\alpha \Delta t}{2 W_g}}
          + \frac{\Delta t}{2}\, [\bU^{\top} \bF_i]_\alpha\,
            \exprel\!\left( -\frac{\kappa_\alpha \Delta t}{2 W_g} \right)
        \right)_{\!\alpha=1,\dots,d},
      \label{eq:action-L2}
  \end{align}
\end{subequations}
where $\kappa_\alpha \coloneqq \rho_\alpha + \Tr[\pg] / N_f$.

\section{Isothermal--isobaric integrator with masked MTK barostat}
\label{app:masked-integrator}

For the masked case with $n_c$ active axes, the Liouville operator splitting [Eq.~\eqref{eq:liouville-decomp}] and the Trotter factorization [Eq.~\eqref{eq:trotter}] remain unchanged.
The only modification is to $i\mathcal{L}_{g,2}$, which now sums only over active axes:
\begin{align}
  i\mathcal{L}_{g,2} \coloneqq \sum_{c=1}^{n_c} G_c \frac{\partial}{\partial p_c},
  \label{eq:Lg2-masked}
\end{align}
with the per-axis driving force
\begin{align}
  G_c \coloneqq \det[\bh] \cdot \be_c^{\top}
    (\mathbf{P}^{\mathrm{int}} - P\bI)\, \be_c
    + \frac{1}{N_f} \sum_{i=1}^{N} \frac{\bp_i^{2}}{m_i}.
  \label{eq:Gc-masked}
\end{align}

Since each barostat momentum $p_c$ is a scalar, the NHC-barostat operator $i\mathcal{L}_{\mathrm{NHC\text{-}baro}}$ acts on each active axis independently.
Its explicit form is analogous to the isotropic case [Eq.~\eqref{eq:nhc-baro-iso}], with $p_\epsilon$ replaced by the sum over active momenta:
\begin{align}
  i\mathcal{L}_{\mathrm{NHC\text{-}baro}}
    &\coloneqq -\frac{p_{\xi_1}}{Q'_1} \sum_{c=1}^{n_c} p_c \frac{\partial}{\partial p_c}
      + \sum_{j=1}^{M} \frac{p_{\xi_j}}{Q'_j} \frac{\partial}{\partial \xi_j}
      + \sum_{j=1}^{M-1}
        \left( G'_j - p_{\xi_j} \frac{p_{\xi_{j+1}}}{Q'_{j+1}} \right)
        \frac{\partial}{\partial p_{\xi_j}}
      + G'_M \frac{\partial}{\partial p_{\xi_M}},
  \label{eq:nhc-baro-masked}
\end{align}
with driving forces $G'_j$ as defined in Eq.~\eqref{eq:masked-driving}.

The masked $\pg$ [Eq.~\eqref{eq:pg-masked}] is already diagonal in the $\be_c$ basis with entries $\rho_c = p_c$ on active axes and $\rho_c = 0$ on inactive axes, so that $\bU = \bI$.
The actions of $e^{i\mathcal{L}_1 \Delta t}$ and $e^{i\mathcal{L}_2 \frac{\Delta t}{2}}$ [Eqs.~\eqref{eq:action-L1}--\eqref{eq:action-L2}] therefore apply directly without requiring a coordinate transformation.

\bibliography{references}

\end{document}